\documentclass[aps,superscriptaddress,reprint]{revtex4-2}
\usepackage{amsmath,amssymb,bm,graphicx,microtype}
\usepackage{hyperref}
\usepackage{braket}
\usepackage{graphicx}
\usepackage{mathrsfs}
\usepackage{xcolor}
\newcommand{\ii}{\mathrm{i}}
\newcommand{\Var}{\operatorname{Var}}
\newcommand{\Cov}{\operatorname{Cov}}

\begin{document}

\title{Programmable Heisenberg-limit sensor from a nonlinear quantum energy pump}

\author{Yang Peng}
\affiliation{Department of Physics and Astronomy, California State University, Northridge, California 91330, USA}
\affiliation{Institute of Quantum Information and Matter and Department of Physics, California Institute of Technology, Pasadena, California 91125, USA}
\email{yang.peng@csun.edu}

\begin{abstract}
We introduce a programmable Heisenberg-limited bosonic quantum sensor based on a nonlinear quantum energy pump, implemented with a Kerr-nonlinear resonator coupled to multiple high-Q microwave terminal resonators. For parameter estimation encoded in an arbitrary number-conserving Hamiltonian acting on the terminal modes, we analytically construct optimal sensing protocols that attain the maximal quantum Fisher information, including initial-state loading, probe preparation, and readout. For diagonal multiparameter signals, we further show that the full phase-sensing quantum Fisher information matrix can be obtained from correlations of locally measured physical terminal works, providing a signal-free calibration of the metrological resource. We numerically demonstrate the construction and its robustness using realistic circuit-QED parameters while including experimentally relevant imperfections.
\end{abstract}

\maketitle
\emph{Introduction.---}
Quantum sensors infer weak fields, frequency shifts, forces, and other physical parameters from the controlled response of a quantum system. They underpin precision measurements ranging from magnetometry and atomic clocks to gravitational-wave detection and searches for new physics \cite{Degen2017,YeZoller2024,Aasi2013}. For an interferometer supplied with \(N\) independent excitations, the quantum Fisher information (QFI) typically scales as \(N\), leading to the standard quantum limit. Entangled or otherwise nonclassical probes can instead achieve a QFI proportional to \(N^2\), known as the Heisenberg-limit scaling \cite{Giovannetti2006,Giovannetti2011,Dowling2008,Pezze2018,Toth2014}. The central challenge is therefore not only to identify probes with Heisenberg-limited sensitivity, but also to determine how such probes should be prepared and measured for the particular Hamiltonian through which the unknown parameter is encoded.

Although the QFI identifies the maximum sensitivity permitted by a given probe and parameter-encoding dynamics, it does not by itself provide an experimental prescription for attaining that sensitivity \cite{Braunstein1994}. For a sensor operated with a fixed number of excitations, the optimal probe depends both on the operator through which the unknown parameter enters and on the known dynamics acting during the interrogation time. Exact local generators and spectral bounds identify the states that maximize the QFI for general Hamiltonian parameters \cite{PangBrun2014,PangJordan2017,Fiderer2019}, and optimal probes have been characterized for general two-mode, number-conserving Hamiltonians \cite{Volkoff2016}. Other approaches optimize the sensing generator for a prescribed state or search variationally for probes adapted to noise and experimental constraints \cite{Reilly2023,Long2026}. A general and experimentally implementable procedure that starts from a given sensing Hamiltonian and specifies how to prepare and measure an optimal probe is nevertheless still lacking.

Bosonic cavities are particularly well suited to realizing such a procedure. A long-lived cavity mode provides a large Hilbert space within a single physical element, while ancillary superconducting circuits enable state preparation, nonlinear control, and number- or parity-resolved measurement \cite{Hofheinz2008,Hofheinz2009,Sun2014}. Experiments have demonstrated Heisenberg-limited single-mode metrology in superconducting resonators \cite{Wang2019}, adaptable phase and displacement sensing with a bosonic mode \cite{Pan2025}, and critical frequency sensing with a parametrically driven nonlinear resonator \cite{Beaulieu2025}. High-fidelity exchange between cavity modes \cite{LuBeamSplitter2023,Chapman2023} and rapid non-Gaussian evolution in tunable nonlinear resonators \cite{YurkeStoler1986,Grimm2020,He2023,Ding2025} provide the complementary controls required to prepare and decode nonclassical multimode probes. Controlled energy transfer can itself be used for state preparation; for example, Chen \emph{et al.} showed that ancilla-dependent Floquet energy currents can generate a path-entangled bosonic sensing state \cite{Chen2025}. These developments motivate the use of a nonlinear quantum energy pump to prepare sensing states in long-lived bosonic cavities. Such a pump can be implemented experimentally using a flux-tunable SNAIL resonator \cite{Frattini2017,Grimm2020,LuKerr2023,He2023}, whose tunable self-Kerr interaction supplies the nonlinear evolution required for non-Gaussian state preparation.

\begin{figure}[t]
\includegraphics[width=0.45\textwidth]{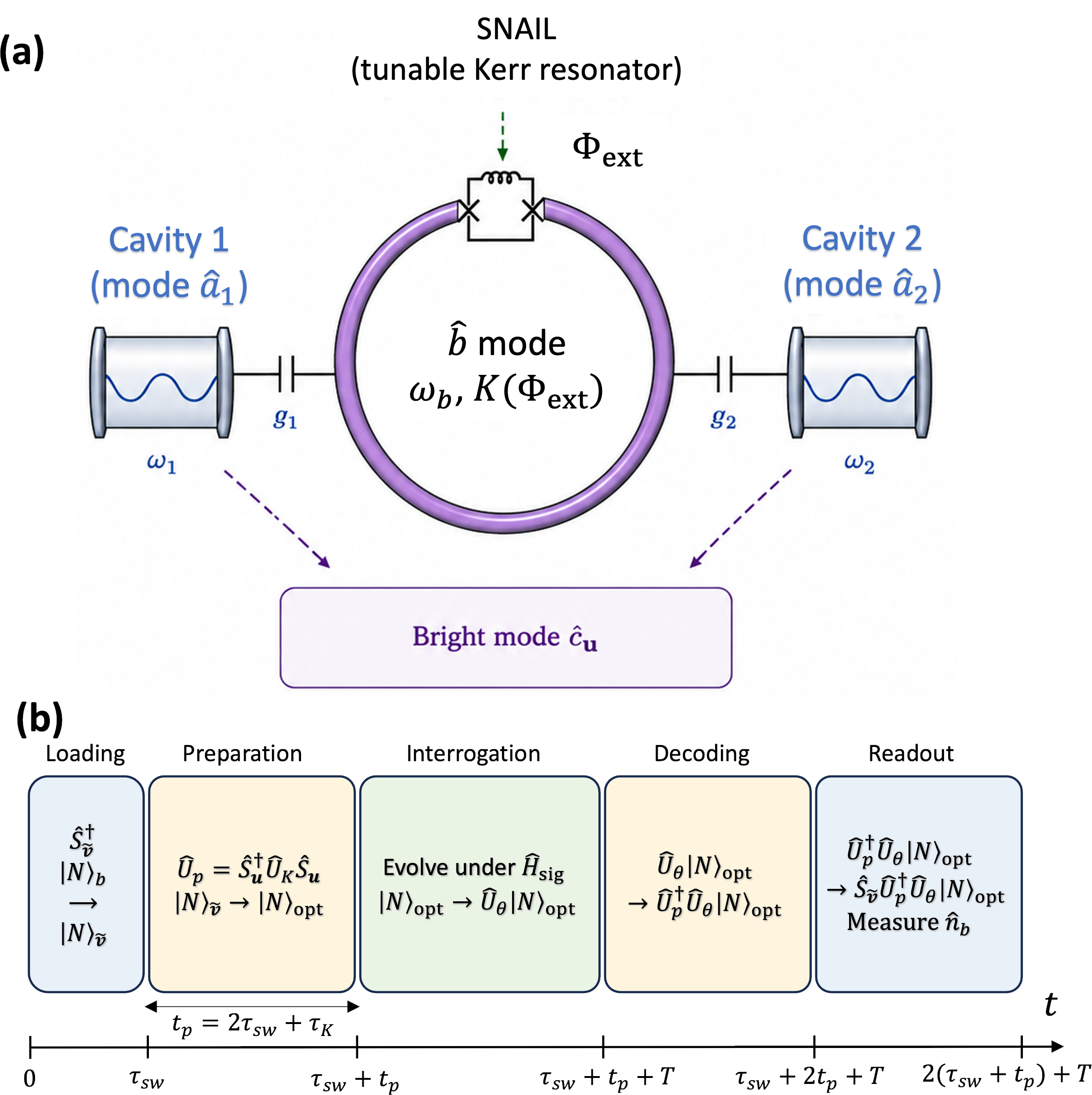}
\caption{
Programmable sensing architecture with circuit-QED implementation.
(a) A flux-tunable SNAIL resonator provides the nonlinear pump mode
\(\hat b\) and is parametrically coupled to two terminal cavities
\(\hat a_1\) and \(\hat a_2\). The relative coupling amplitudes and phases
select the collective bright mode \(\hat c_{\bm u}\) that couples to the
pump, while the external flux controls the pump Kerr interaction.
(b) Pulse sequence for initial state loading, preparation, sensing, decoding, and readout.
A complete swap \(\hat S_{\tilde{\bm v}}^\dagger\) loads the initial pump
Fock state \(\ket{N}_b\) into the terminal mode
\(\ket{N}_{\tilde{\bm v}}\). The nonlinear preparation
\(\hat U_p=\hat S_{\bm u}^\dagger\hat U_K\hat S_{\bm u}\) generates the
optimal probe $\ket{N}_{\rm opt}$, which evolves under the sensing Hamiltonian for time \(T\).
The same nonlinear sequence decodes the sensed state, and
\(\hat S_{\tilde{\bm v}}\) maps the return amplitude back to the pump for
the final number-resolved measurement of \(\hat n_b\).
}
\label{fig:concept}
\end{figure}

Here we introduce a programmable bosonic sensor that provides a constructive solution to the probe-design problem. The sensor consists of several bosonic terminals coupled with independently controlled amplitudes and phases to a nonlinear quantum energy pump~\cite{Peng2026Work,Pengnonabelian,Schmid2026}, as illustrated in Fig.~\ref{fig:concept}(a) for two terminals. For any Hermitian, number-conserving one-body sensing Hamiltonian and a chosen interrogation time, we determine the pump couplings that prepare the optimal fixed-number probe and specify an inverse sequence that reads out the encoded parameter. The resulting probe attains the maximum QFI allowed by the exact finite-time sensing dynamics and therefore reaches Heisenberg-limit scaling. 

We describe a physical implementation using the self-Kerr nonlinearity of a SNAIL resonator \cite{Frattini2017,Grimm2020,LuKerr2023,He2023,Ding2025}, although the sensing principle applies more generally to any nonlinear element capable of producing the required number-dependent phase evolution. As a secondary result, we show that when considering multiple diagonal signals in the physical terminal basis, the corresponding QFI matrix can be reconstructed from correlations of locally measured terminal-energy changes. These work measurements are not required to estimate the unknown parameter, but provide a signal-free means of characterizing and calibrating the prepared sensor. This establishes a connection between the work statistics of quantum energy pump~\cite{Peng2026Work}, and the performance of the corresponding quantum sensor. The resulting protocol differs from adaptable bosonic sensors designed for a fixed encoding \cite{Pan2025}, energy-current generation of path entanglement \cite{Chen2025}, and numerical searches for optimal probes \cite{Long2026} by providing an explicit map from the sensing Hamiltonian to the required experimental controls.

\emph{Sensing protocol.---}
We now formulate the sensing problem and introduce the pump--terminal architecture that realizes the optimal probe. As illustrated in Fig.~\ref{fig:concept}(a), the device contains \(M\) bosonic terminal modes, with annihilation operators
\(\hat{\bm a}=(\hat a_1,\ldots,\hat a_M)^T\), coupled to a nonlinear pump mode \(\hat b\). The terminal frequencies are \(\Omega_i\), and the pump frequency is \(\omega_b\). The laboratory-frame Hamiltonian is
$\hat H_{\rm lab}(t;\theta) = \hat H_0 + \hat H_{\rm nl}(t)  + \hat H_{\rm cpl}(t) + \hat H_{\rm sig}(t;\theta)$.
The terms $\hat H_{\rm nl}(t)$, $\hat H_{\rm cpl}(t)$, and $\hat H_{\rm sig}(t;\theta)$ will be switched on and off during the different stages of the protocol, determined by the time $t$.

The uncoupled Hamiltonian is $\hat H_0 = \sum_{i=1}^{M}\Omega_i\hat n_i + \omega_b\hat n_b$, with number operators  $\hat n_i=\hat a_i^\dagger\hat a_i$, and  $\hat n_b=\hat b^\dagger\hat b$.
For the SNAIL implementation considered below, the pump nonlinearity is a tunable self-Kerr interaction~\cite{Grimm2020,LuKerr2023,He2023,Ding2025} $\hat H_{\rm nl}(t) = K(t)\, \hat n_b(\hat n_b-1)/2$. Each terminal is coupled parametrically to the pump, via terms $\bigl(\hat a_i+\hat a_i^\dagger\bigr) \bigl(\hat b+\hat b^\dagger\bigr)$, through a separately controlled modulation, see Supplemental Material (SM)~\cite{SM} for an explicit expression).

We describe the unknown parameter \(\theta\) to be estimated by the most general Hermitian, number-conserving one-body signal Hamiltonian acting on the terminals, $\hat H_{\rm sig}(t;\theta)=\hat{\bm a}^{\dagger} Q_\theta(t)\hat{\bm a}$, with
$Q_\theta(t)=Q_\theta^\dagger(t)$.
The diagonal entries of \(Q_\theta\) describe terminal-frequency shifts, whereas its off-diagonal entries describe parameter-dependent couplings between terminals.

To separate the fast bare oscillations from the controlled dynamics, we use the parameter-independent rotating-frame transformation $\hat R(t)=e^{-\ii t\hat H_0}$, which gives the rotated Hamiltonian $\hat H^{(r)} = \hat R^\dagger\hat H_{\rm lab}\hat R - \ii\hat R^\dagger\dot{\hat R}$. In this rotated frame, 
$\hat H^{(r)}(t;\theta)
    =
    \hat H_{\rm nl}(t)
    +
    \hat H_{\rm sw}(t)
    +
    \hat H_{\rm sig}^{(r)}(t;\theta)$.

The pump-terminal coupling under the rotating wave approximation becomes~\cite{LuBeamSplitter2023,Chapman2023,SM}
\begin{equation}
    \hat H_{\rm sw}(t)
    \simeq
    \sum_{i=1}^{M}
    G_i(t)
    \left(
        e^{\ii\phi_i}\hat a_i^\dagger\hat b
        +
        e^{-\ii\phi_i}\hat b^\dagger\hat a_i
    \right),
    \label{eq:Hsw}
\end{equation}
where $G_i(t)$, $\phi_i$  are the effective exchange rates and phases, controlled independently for each terminal. 

We make no rotating-wave approximation to the signal Hamiltonian. Its exact rotating-frame form is
$\hat H_{\rm sig}^{(r)}(t;\theta) = \hat{\bm a}^{\dagger} Q_\theta^{(r)}(t) \hat{\bm a}$, where
$Q_\theta^{(r)}(t) = e^{\ii D_\Omega t} Q_\theta(t) e^{-\ii D_\Omega t}$,
and the diagonal matrix $D_\Omega = \operatorname{diag}(\Omega_1,\ldots,\Omega_M)$ consists of all terminal frequencies.
For a time-independent laboratory-frame matrix \(Q_\theta\), its rotating-frame elements are
$\bigl[Q_\theta^{(r)}(t)\bigr]_{ij} = (Q_\theta)_{ij} e^{\ii(\Omega_i-\Omega_j)t}$.
The diagonal terms therefore remain stationary, while the off-diagonal terms acquire known oscillating phases. We retain all of these terms exactly during interrogation. 

During probe preparation and readout, \(\hat H_{\rm sw}\) and \(\hat H_{\rm nl}\) are applied as controlled pulses. During interrogation, these controls are turned off and the terminals evolve under the exact signal Hamiltonian \(\hat H_{\rm sig}^{(r)}(t;\theta)\). 

\emph{Probe preparation.---}
As illustrated in Fig.~\ref{fig:concept}(b), at the loading state, an $N$-excitation Fock state $\ket{N}_b$ was initialized in the pump and then transferred to the terminals via $\hat{S}_{\tilde{\bm v}}^\dagger$, which will be defined later. 
This is followed by the preparation sequence, 
described by unitary $\hat{U}_p$ consists of a swap pulse $\hat{S}_{\bm u}$, a Kerr
pulse $\hat{U}_K$, and the inverse swap $\hat{S}^\dagger_{\bm u}$.

Let us now define the swap operation $\hat{S}_{\bm u}$ in detail.
During the swap pulse, the coupling amplitudes are held at fixed relative
values. Eq.~(\ref{eq:Hsw}) can then be written as
\begin{equation}
\hat H_{\rm sw}(t)
\simeq 
G(t)\left(
\hat c_{\bm u}^\dagger\hat b
+
\hat b^\dagger\hat c_{\bm u}
\right),
\qquad
\hat c_{\bm u}=\bm u^\dagger\hat{\bm a},
\label{eq:brightMode}
\end{equation}
where
$G=\sqrt{\sum_iG_i^2}$,
and $\bm u\in \mathbb{C}^M$ is normalized with elements
$u_i=G_i e^{\ii\phi_i}/G$.
A swap pulse with fixed \(\bm u\) and pulse area
$\int_0^{\tau_{\rm sw}}G(t)\,dt=\frac{\pi}{2}$ gives a complete 
swap operation $\hat{S}_{\bm u} = \exp[-i\pi(\hat c_{\bm u}^\dagger\hat b +\hat b^\dagger\hat c_{\bm u})/2]$, which
transforms
\(\hat{S}_{\bm u}^\dagger \hat c_{\bm u} \hat{S}_{\bm u}=-\ii\hat b\) and
\(\hat{S}_{\bm u}^\dagger \hat b \hat{S}_{\bm u}=-\ii\hat c_{\bm u}\). 
For constant \(G\), the swap duration is
\(\tau_{\rm sw}=\pi/(2G)\). 
Thus, $\hat{S}_{\bm u}$ (and $\hat{S}_{\bm u}^\dagger$) exchanges the $N$ excitations between
the pump and the terminal superposition state (here is the bright state) specified by $\bm u$, 
namely $\hat{S}_{\bm u}\ket{N}_{\bm u}\ket{0}_b \sim \ket{0}_{\bm u}\ket{N}_b$, up to a phase.
The dark terminal modes that are orthogonal to the bright state remain unchanged. Here, we use the notation $\ket{N}_{\bm u} = (\hat{c}_{\bm u}^\dagger)^N \ket{0}_t/\sqrt{N!}$, which has $N$ excitations carried by the mode $\bm u$, and $\ket{0}_t = \ket{0}_{\bm v}$ for any $\bm v$ denotes the terminal vacuum state.

During the Kerr pulse, a constant Kerr strength \(K>0\) is switched on, and the system evolves under
\(\hat H_{\rm nl}=K\hat n_b(\hat n_b-1)/2\) for a duration
\(\tau_K=\pi/K\). The corresponding evolution operator is
$\hat{U}_K = \exp\!\left[ -\ii\frac{\pi}{2}\hat n_b(\hat n_b-1) \right]$.
Since $\hat{S}_{\bm u}^\dagger \hat{n}_b \hat{S}_{\bm u} = \hat{c}_{\bm u}^\dagger\hat{c}_{\bm u} \equiv \hat{n}_{\bm u}$, we find the full preparation sequence can be written as~\cite{SM}
\begin{equation}
    \hat U_p = \hat{S}_{\bm u}^\dagger \hat{U}_K \hat{S}_{\bm u}
    = \frac{e^{-i\pi/4}}{\sqrt 2}
    \left[e^{i\frac{\pi}{2}\hat{n}_{\bm u}} +i e^{-i\frac{\pi}{2}\hat{n}_{\bm u}} \right].
\end{equation}

For a pump initially in \(\ket{0}_b\), the complete preparation sequence returns the pump to vacuum and induces the terminal unitary $\hat{U}_p$. To see its action on the terminal states, we consider a generic $N$-excitation terminal state $\ket{N}_{\bm{v}}$, where $\bm v \in \mathbb{C}^M$ is an arbitrary normalized vector. We find
\begin{equation}
\hat{U}_p \ket{N}_{\bm v} = \frac{e^{-i\pi/4}}{\sqrt{2}}\left[\ket{N}_{V_{\bm u}(\frac{\pi}{2})\bm v} + i\ket{N}_{V_{\bm u}(-\frac{\pi}{2})\bm v} \right],
\label{eq:Uptransform}
\end{equation}
where we introduced the phase rotation matrix
$V_{\bm u}(\alpha) = (\mathbb{I} - {\bm u}{\bm u}^\dagger) + e^{i\alpha}{\bm u}{\bm u}^\dagger$,
where $\mathbb{I}$ is the identity operator. 

In the following, we show that the vectors $\tilde{\bm v}$, $\bm u$ for the loading and preparation, can be chosen depending on the signal Hamiltonian, such that the maximal sensitivity can be achieved. 

\emph{Programmable optimal probe.---}
We consider local estimation around a chosen operating point \(\theta_0\).  After preparation, the pump controls are turned off and the terminals evolve under \(\hat H_{\rm sig}^{(r)}(t;\theta)\) for an interrogation time \(T\). 
Its time evolution operator is
$\hat{U}_\theta(T) =\mathcal{T}\exp\left[-\ii\int_0^T dt\, \hat H_{\rm sig}^{(r)}(t;\theta)\right]$, where $\mathcal{T}$ denotes the time ordering.
We introduce the Hermitian matrix \(K_{\theta_0}(T)\) describes the parameter dependence accumulated over this finite interval, which can be obtained via $\hat{\bm a}^\dagger K_{\theta_0}(T) \hat{\bm a} =i \left. \hat{U}^\dagger_{\theta_0}(T)\partial_\theta \hat{U}_{\theta}(T)\right |_{\theta = \theta_0}$~\cite{PangBrun2014,PangJordan2017}.

For a pure input probe state $\ket{\psi}$, the QFI at \(\theta_0\) is
$F_Q(\theta_0) = 4\,\operatorname{Var}_{\psi}
\!\left[
\hat{\bm a}^{\dagger}
K_{\theta_0}(T)
\hat{\bm a}
\right]$.
Thus, the complete time-dependent interrogation enters the local
sensitivity through the finite-time generator \(K_{\theta_0}(T)\).
Within the fixed-\(N\) sector, the largest and smallest eigenvalues of
\(\hat{\bm a}^{\dagger}K_{\theta_0}(T)\hat{\bm a}\)
are \(N\kappa_+\) and \(N\kappa_-\), respectively. The largest possible
variance is therefore one quarter of the squared spectral range~\cite{BhatiaDavis2000}, giving
$F_Q^{\max}(T)
=
N^2\bigl(\kappa_+-\kappa_-\bigr)^2$.
The bound is attained by the balanced extremal-mode superposition
\begin{equation}
\ket{N}_{\rm opt}
=
\frac{
\ket{N}_{\bm v_-}
+
e^{\ii\gamma}\ket{N}_{\bm v_+}
}{\sqrt{2}},
\label{eq:optimalProbe}
\end{equation}
where the extra phase \(\gamma\) does not affect the QFI
\cite{PangBrun2014,Fiderer2019}. 

To deterministically prepare the probe that reach the abstract optimum specified by Eq.~\eqref{eq:optimalProbe}, we choose the complex vector 
$\bm u = (\bm v_- - e^{i \chi} \bm v_+)/\sqrt{2}$ in the preparation stage,
which encodes the couplings between terminals and the pump, where $\chi$ can be an arbitrary phase. For the initial terminal state, we require it be $\ket{N}_{\tilde{\bm v}}$, where the vector $\tilde{\bm v} = V_{\bm u}(-\pi/2){\bm v}_-$. This state is obtained from an $N$-excitation pump state~\cite{Hofheinz2008,Hofheinz2009}, under the action of the complete swap operator $\hat{S}_{\tilde{\bm v}}^\dagger$ on $\ket{0}_t\ket{N}_b$.

Straightforward calculation using shows that
\begin{equation}
    \hat{U}_p \ket{N}_{\tilde{\bm v}} = 
    \frac{e^{-i\frac{\pi}{4}}}{\sqrt{2}}
    \left[\ket{N}_{\bm v_-} +i e^{iN\chi} \ket{N}_{\bm v_+}  \right] \sim \ket{N_{\rm opt}}.
\end{equation}

\emph{Readout.---}
The sensing signal is converted into a return probability using the same
pulse sequences employed for probe preparation~\cite{Macri2016,Nolan2017,Haine2018,Colombo2022}. After interrogation, we
first apply a calibrated analysis operation corresponding to the inverse
evolution at the operating point,
\(\hat U_{\theta_0}^{\dagger}(T)\). The resulting terminal state is $\hat U_{\theta_0}^{\dagger}(T)
\hat U_{\theta}(T) \ket{N}_{\rm opt}$. At \(\theta=\theta_0\), this state is exactly
\(\ket{N}_{\rm opt}\), whereas a deviation \(\theta-\theta_0\) produces a corresponding deviation from the optimal
probe.

We then apply the same swap--Kerr--inverse-swap pulse sequence
\(\hat U_p\) used for preparation. Since $\hat U_p^2
= \left(
\hat S_{\bm u}^{\dagger}
\hat U_K
\hat S_{\bm u}
\right)^2
= \mathbb{I}$, at the operating point it reverses the preparation, via
$\ket{N}_{\rm opt}
\to
\ket{N}_{\tilde{\bm v}}$,
up to a known phase. For \(\theta\neq\theta_0\), however,
\(\hat U_{\theta_0}^{\dagger}(T)\hat U_{\theta}(T)
\ket{N}_{\rm opt}\neq\ket{N}_{\rm opt}\), and the same pulse
sequence therefore does not return the terminals perfectly to
\(\ket{N_{\tilde{\bm v}}}\). This imperfect return carries the sensing
signal.

Finally, the inverse loading swap transfers the component in
\(\ket{N}_{\tilde{\bm v}}\) back to the pump, via $\hat{S}_{\tilde{\bm v}}\ket{N}_{\tilde{\bm v}}\ket{0}_b \sim\ket{0}_t\ket{N}_b$,
so that the readout reduces to detecting whether the pump contains exactly
\(N\) excitations~\cite{Sun2014,LuKerr2023}. The probability of detecting \(N\) excitations in the pump is therefore
$p_N(\theta)
=
\left|
{}_{\rm opt}\bra{N}
\hat U_{\theta_0}^{\dagger}(T)
\hat U_{\theta}(T)
\ket{N}_{\rm opt}
\right|^2$
Repeating the
protocol and recording the binary outcomes \(n_b=N\) or \(n_b\neq N\)
locally attains the optimal Fisher information (FI).


\emph{Physical-work certification for diagonal signals.---}
A particularly transparent connection to physical work arises for diagonal
multiparameter sensing~\cite{Humphreys2013,Ragy2016}. Consider \(\hat H_{\rm sig}(\bm\theta)=\sum_{i=1}^{M}\theta_i\hat n_i\), where
\(\theta_i\) is the unknown frequency shift of terminal \(i\). After an
interrogation time \(T\), the generators of the parameters \(\theta_i\) are
\(T\hat n_i\).

These number correlations can be determined from the physical energy
transferred to the terminals during probe preparation. Defining the work
deposited in terminal \(i\) as
\(W_i=\Omega_i(\hat U_p^\dagger\hat n_i\hat U_p-\hat n_i)\)~\cite{Peng2026Work} and assuming
sharp terminal occupations before preparation, one obtains the QFI matrix of a pure state probe
\begin{equation}
(F_Q)_{ij}
=4T^2\,{\rm Cov}(\hat n_i,\hat n_j) = 
\frac{4T^2}{\Omega_i\Omega_j}\,
{\rm Cov}(W_i,W_j).
\label{eq:workQFI}
\end{equation}

Thus, for diagonal signals, the full QFI matrix can be certified from
correlations of the terminal-energy changes generated during preparation,
without applying the unknown signal. This provides a signal-free
calibration of the metrological resource.

\emph{Two-terminal examples.---}
We illustrate the protocol with two terminals. First consider a
single-parameter diagonal signal
\(\hat H_{\rm sig}(\lambda)
=\lambda(q_1\hat n_1+q_2\hat n_2)\), so that the full laboratory-frame
Hamiltonian during interrogation is
\(\hat H_0+\hat H_{\rm sig}(\lambda)\), with
\(\hat H_0=\Omega_1\hat n_1+\Omega_2\hat n_2\).
Since the signal commutes with \(\hat H_0\), it is unchanged in the rotating
frame and
\(K_{\lambda_0}(T)=T\,{\rm diag}(q_1,q_2)\).
For \(q_1>q_2\), the extremal modes are simply the two physical terminals,
and the optimal bright mode therefore requires equal-magnitude coupling to
them. The maximal QFI is
$F_Q^{\max}
=
N^2T^2(q_1-q_2)^2$.
At \((q_1,q_2)=(1/2,-1/2)\), the ideal return measurement gives
\(p_N=\cos^2(N\lambda T/2)\), and hence the classical FI $F_C$ saturates to $F_Q^{\max}$.
This single-parameter example corresponds to sensing one direction
\(\bm q=(q_1,q_2)\) of the diagonal multiparameter problem discussed above.

As a more stringent example, consider the off-diagonal signal
$\hat H_{\rm sig}(\theta)
=
\frac{\theta}{2}
\left(
\hat a_1^\dagger\hat a_2
+
\hat a_2^\dagger\hat a_1
\right)$.
Unlike the diagonal case, this signal does not commute with
\(\hat H_0\). At the operating point \(\theta_0=0\), defining
\(\Delta_\Omega=\Omega_1-\Omega_2\), the exact finite-time generator is
\begin{equation}
K_{0}(T)
=
\frac{T}{2}\,
{\rm sinc}\!\left(\frac{\Delta_\Omega T}{2}\right)
\begin{pmatrix}
0 & e^{\ii\Delta_\Omega T/2}\\
e^{-\ii\Delta_\Omega T/2} & 0
\end{pmatrix},
\label{eq:offdiagKT}
\end{equation}
which gives
$F_Q^{\max}
=
N^2T^2\,
{\rm sinc}^2\!\left(\frac{\Delta_\Omega T}{2}\right)$,
with ${\rm sinc}(x) = \sin(x)/x$.
Thus the terminal-frequency mismatch produces a finite-time filtering of
the coherent-transfer signal.  The required bright mode follows
directly from the extremal eigenvectors of $K_0(T)$. A
three-terminal example, together with control imperfections, is
given in the SM~\cite{SM}.

\begin{figure}[t]
\includegraphics[width=\columnwidth]{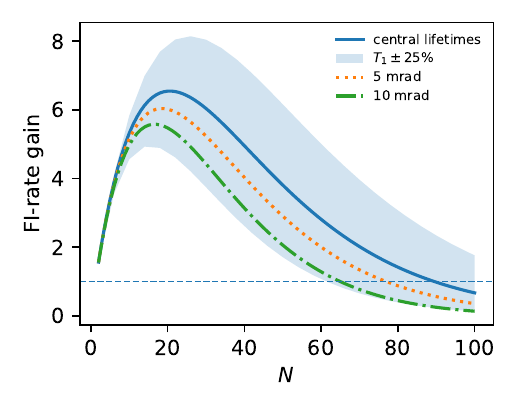}
\caption{Realistic two-terminal benchmark for the differential signal
\(\hat H_{\rm sig}=\lambda(\hat n_1-\hat n_2)/2\) using the complete
loading--preparation--interrogation--decoding--unloading protocol.
Shown is the FI-rate gain \(\mathcal R\) relative to a
lossy separable sensor with zero control overhead. For
\(G/2\pi=2.05\,{\rm MHz}\), \(|K|/2\pi=5.21\,{\rm MHz}\),
\(T_{1,T}=204\,\mu{\rm s}\), \(T_{1,P}=16\,\mu{\rm s}\), and
\(T=5\,\mu{\rm s}\), the gain peaks at \(\mathcal R\simeq6.55\) near
\(N=21\). The shaded band varies both lifetimes by \(\pm25\%\); dotted
and dash-dotted curves include quasistatic differential-phase noise of
\(5\) and \(10\) mrad rms, respectively. The blue dashed line indicates $\mathcal{R}=1$.}
\label{fig:numerics}
\end{figure}

\emph{Experimental realization.---}
The protocol can be implemented in circuit QED using high-\(Q\) microwave
terminal resonators coupled to a flux-tunable SNAIL resonator that acts as
the nonlinear pump. The SNAIL provides the tunable self-Kerr interaction,
while parametric modulation of each terminal--pump coupling realizes the
programmable exchange in Eq.~\eqref{eq:Hsw}
\cite{Frattini2017,LuKerr2023,He2023,LuBeamSplitter2023}.
During the swap pulses the SNAIL is biased near its Kerr-free point, and
during the nonlinear pulse it is rapidly tuned to a finite Kerr strength.

The rotating-wave approximation is required only for these engineered
exchange pulses. It is valid when the exchange rates and pulse bandwidths
are small compared with the bare resonator frequencies and the frequencies
of the counter-rotating processes. With GHz-scale resonators and
MHz-scale exchange rates this condition is well satisfied.

For a realistic benchmark, we consider the two-terminal differential signal
\(\hat H_{\rm sig}=\lambda(\hat n_1-\hat n_2)/2\).  In this case
\(\bm u=(\bm e_2-\bm e_1)/\sqrt2\) and
\(\tilde{\bm v}\sim(\bm e_1-\ii\bm e_2)/\sqrt2\), where $\bm e_i$ is the standard basis of $\mathbb{R}^2$.  We take
\(G/2\pi=2.05\,{\rm MHz}\), \(|K|/2\pi=5.21\,{\rm MHz}\),
terminal lifetime \(T_{1,T}=204\,\mu{\rm s}\), pump lifetime
\(T_{1,P}=16\,\mu{\rm s}\), and interrogation time
\(T=5\,\mu{\rm s}\).  The complete-swap and Kerr-pulse durations are
\(\tau_{\rm sw}=\pi/(2G)=0.122\,\mu{\rm s}\) and
\(\tau_K=\pi/|K|=0.096\,\mu{\rm s}\), respectively.  Thus the loading and preparation
(or decoding and readout) sequence takes
\(3\tau_{\rm sw}+\tau_K=0.462\,\mu{\rm s}\), much smaller than the lifetimes.

We numerically propagate the complete
loading--preparation--interrogation--decoding--readout sequence in the
fixed-total-$N$-excitation basis, whose dimension is
\((N+1)(N+2)/2\)~\cite{SM}.  The swap pulses are evaluated by sparse matrix
exponentiation, while the Kerr pulse and the differential signal are
diagonal in the Fock basis.  Photon loss is included throughout the
sequence by replacing the instantaneous Hamiltonian by a non-hermitian Hamiltonian
$
\hat H(t)-\frac{\ii}{2}\left[
(\hat n_1+\hat n_2)/T_{1,T}
+\hat n_b/T_{1,P}
\right].
$
For the measured return event \(n_b=N\), this non-Hermitian evolution is
exact: any photon-loss jump lowers the total excitation number, and the
subsequent number-conserving controls cannot return that trajectory to
\(n_b=N\).  The resulting return probability \(p_N\) therefore directly
gives the binary FI
\(F_C=(\partial p_N)^2/[p_N(1-p_N)]\), maximized over $\lambda \in(0,\pi/NT]$.

To include quasistatic phase noise, we first average \(p_N\) over a
zero-mean Gaussian fluctuation of the accumulated differential phase and
then evaluate \(F_C\). 
To compare sensing speed, we define the \emph{FI-rate gain}
\(\mathcal R=\tilde F_C/(N\tilde F_{\rm sep})\), where
\(\tilde F_C=F_C/[2(3\tau_{\rm sw}+\tau_K)+T]\) is the FI
per unit time of the nonlinear protocol, including preparation and
decoding, while
\(\tilde F_{\rm sep}=e^{-T/T_{1,T}}/T\) is that of a single-photon probe
with terminal loss and no control overhead. Quasistatic phase noise is
included by averaging \(p_N\) over a zero-mean Gaussian differential-phase
fluctuation before evaluating \(F_C\).
Thus \(\mathcal R>1\) means that the nonlinear sensor acquires more
information per unit experimental time than the lossy separable sensor,
even after including its preparation and decoding overhead. 

Figure~\ref{fig:numerics} shows that the central-lifetime gain reaches
\(\mathcal R\simeq6.55\) at \(N=21\).  The shaded band varies both
lifetimes by \(\pm25\%\), while the dotted and dash-dotted curves include
\(5\) and \(10\) mrad rms quasistatic differential phase noise,
respectively.  The optimum at finite \(N\) reflects the competition
between the \(N^2\) coherent enhancement and the increasing loss
sensitivity of larger Fock-state superpositions~\cite{Huelga1997,Escher2011,Demkowicz2009,Demkowicz2012}.  

\emph{Conclusion.---}
We have presented a constructive route from an arbitrary Hermitian,
number-conserving one-body sensing Hamiltonian to a Heisenberg-limited
bosonic sensor. The initial state and preparation sequence that generate
the optimal probe attaining the maximal QFI can be programmed through the
pump--terminal couplings and the nonlinear Kerr interaction of the pump.
The same pulse sequence also enables a number-resolved return measurement
through the pump. For diagonal multiparameter signals, the corresponding
QFI matrix can additionally be certified from physical terminal-energy
correlations, providing a signal-free calibration of the metrological
resource.

The proposed sensor can be implemented in circuit QED with a
SNAIL resonator as the nonlinear pump. In a two-terminal example with
realistic parameters, our numerics shows that the metrological rate
advantage survives finite control times, photon loss, and phase noise.
This establishes nonlinear quantum energy pumps as programmable platform
for quantum metrology.

Several extensions are natural. An important direction is to determine
optimal preparation and readout strategies in the presence of strong loss
and dephasing, where the optimal probe need not coincide with its
closed-system counterpart. It will also be interesting to extend the
programmable construction to genuinely multiparameter noncommuting signals
and to larger bosonic networks, where the pump may provide a scalable
route to tailoring many-body metrological resources to increasingly
complex sensing Hamiltonians.

\emph{Acknowledgment.}--- This work is supported by the US National Science Foundation (NSF) Grant No. PHY-2216774.
The numerical simulation is supported by NSF instrument grant  DMR-2406524.
%

\newpage

\begin{widetext}
\section*{Supplemental material}
This Supplemental Material derives the rotating-frame control Hamiltonians, the exact finite-time generator of the sensing dynamics, the fixed-number spectral bound, the programmable preparation and readout, numerical details of two-terminal and three-terminal simulations, and the physical terminal-work correspondence for diagonal signals.

\section{Rotating-frame controls}
\label{S:rotatingControls}

A convenient laboratory-frame realization of the pump--terminal coupling is
\begin{equation}
 \hat H_{\rm cpl}(t)
 =\sum_i2g_i(t)\cos[\nu_i t+\phi_i]
 (\hat a_i+\hat a_i^{\dagger})(\hat b+\hat b^{\dagger}).
 \label{S:labCoupling}
\end{equation}
Choosing \(\nu_i=|\omega_b-\Omega_i|\) and transforming with
\(\hat R(t)=e^{-\ii t\hat H_0}\) leaves the resonant exchange term
\begin{equation}
 \hat H_{\rm sw}(t)
 \simeq\sum_iG_i(t)
 \left(e^{\ii\phi_i}\hat a_i^{\dagger}\hat b
 +e^{-\ii\phi_i}\hat b^{\dagger}\hat a_i\right),
\end{equation}
where the slowly varying amplitude is absorbed into \(G_i(t)\). The discarded control terms rotate at frequencies of order
\(2|\omega_b-\Omega_i|\), \(2\Omega_i\), or \(2\omega_b\). A sufficient condition over the occupied manifold is therefore
\begin{equation}
 NG_i,\ \tau_{\rm edge}^{-1},\ |\delta_i|
 \ll
 \min\{2|\omega_b-\Omega_i|,2\Omega_i,2\omega_b\},
 \label{S:RWAcondition}
\end{equation}
where \(\tau_{\rm edge}\) characterizes the pulse turn-on/off and \(\delta_i\) is a residual modulation detuning. During swaps, residual pump Kerr should additionally satisfy
\begin{equation}
 |K|(N-1)\ll G.
\end{equation}
This is why the SNAIL is biased near its Kerr-free point during exchange pulses and moved to finite \(K\) only for the nonlinear step.

The rotating-wave approximation above is made only for the engineered controls. The sensing term is transformed exactly,
\begin{equation}
 \hat H_{\rm sig}^{(r)}(t;\theta)
 =\hat{\bm a}^{\dagger}Q_{\theta}^{(r)}(t)\hat{\bm a},
 \qquad
 Q_{\theta}^{(r)}(t)
 =e^{\ii D_\Omega t}Q_\theta(t)e^{-\ii D_\Omega t},
\end{equation}
with \(D_\Omega={\rm diag}(\Omega_1,\ldots,\Omega_M)\). For a static laboratory-frame matrix,
\begin{equation}
 [Q_{\theta}^{(r)}(t)]_{ij}
 =(Q_\theta)_{ij}e^{\ii(\Omega_i-\Omega_j)t}.
\end{equation}
These phases are retained in the time-ordered interrogation propagator and are responsible, for example, for the sinc filtering in the off-diagonal two-terminal example in the main text.

\section{Exact finite-time generator and fixed-number bound}
\label{S:generator}

During interrogation the terminals evolve under the number-conserving one-body Hamiltonian
\begin{equation}
 \hat H_{\rm sig}^{(r)}(t;\theta)
 =\hat{\bm a}^{\dagger}Q_{\theta}^{(r)}(t)\hat{\bm a},
 \qquad Q_{\theta}^{(r)}(t)=Q_{\theta}^{(r)\dagger}(t),
\end{equation}
with exact propagator
\begin{equation}
 \hat U_{\theta}(T)=\mathcal T
 \exp\!\left[-\ii\int_0^Tdt\,\hat H_{\rm sig}^{(r)}(t;\theta)\right].
\end{equation}
For local estimation around \(\theta_0\), define
\begin{equation}
 \hat{\mathcal G}_{\theta_0}(T)
 =\ii\hat U_{\theta_0}^{\dagger}(T)
 \left.\partial_{\theta}\hat U_{\theta}(T)\right|_{\theta_0}.
 \label{S:localGenerator}
\end{equation}
Because the dynamics is one-body and number conserving,
\begin{equation}
 \hat{\mathcal G}_{\theta_0}(T)
 =\hat{\bm a}^{\dagger}K_{\theta_0}(T)\hat{\bm a},
 \qquad K_{\theta_0}=K_{\theta_0}^{\dagger}.
\end{equation}
Equivalently, if \(u_{\theta_0}(t)\) is the single-particle propagator generated by \(Q_{\theta_0}^{(r)}(t)\), then
\begin{equation}
 K_{\theta_0}(T)
 =\int_0^Tdt\,
 u_{\theta_0}^{\dagger}(t)
 \left.\partial_{\theta}Q_{\theta}^{(r)}(t)\right|_{\theta_0}
 u_{\theta_0}(t).
 \label{S:Kintegral}
\end{equation}
This expression keeps the complete finite-time sensing dynamics and requires no secular approximation to the signal.

Let \(\kappa_-\) and \(\kappa_+\) be the smallest and largest eigenvalues of \(K_{\theta_0}(T)\), with normalized eigenvectors \(\bm v_-\) and \(\bm v_+\). In the sector containing exactly \(N\) terminal excitations, the many-body generator has extremal eigenvalues \(N\kappa_-\) and \(N\kappa_+\), attained by
\begin{equation}
 \ket{N}_{\bm v}
 =\frac{(\hat{\bm a}^{\dagger}\bm v)^N}{\sqrt{N!}}\ket{0}_t.
\end{equation}
For a pure probe, \(F_Q=4\Var(\hat{\mathcal G}_{\theta_0})\). The largest variance of a Hermitian operator is one quarter of the squared spectral range; hence
\begin{equation}
 F_Q^{\max}
 =N^2(\kappa_+-\kappa_-)^2,
 \label{S:fixedNbound}
\end{equation}
attained by the equal-weight extremal superposition
\begin{equation}
 \ket{N_{\rm opt}}
 =\frac{\ket{N}_{\bm v_-}+e^{\ii\gamma}\ket{N}_{\bm v_+}}{\sqrt2}.
 \label{S:optimalCat}
\end{equation}
The phase \(\gamma\) does not affect the QFI. Convexity of the QFI implies that allowing mixed states cannot increase the fixed-\(N\) optimum.

\section{Programmable preparation and readout}
\label{S:programming}

For a normalized terminal vector \(\bm u\), define the bright mode
\(\hat c_{\bm u}=\bm u^{\dagger}\hat{\bm a}\). A resonant exchange pulse implements
\begin{equation}
 \hat H_{\rm sw}=G(t)
 \left(\hat c_{\bm u}^{\dagger}\hat b+\hat b^{\dagger}\hat c_{\bm u}\right),
 \qquad
 \int dt\,G(t)=\frac{\pi}{2},
\end{equation}
and therefore the complete swap
\begin{equation}
 \hat S_{\bm u}
 =\exp\!\left[-\ii\frac{\pi}{2}
 \left(\hat c_{\bm u}^{\dagger}\hat b+\hat b^{\dagger}\hat c_{\bm u}\right)\right].
\end{equation}
For the Kerr pulse area used in the main text,
\begin{equation}
 \hat U_K
 =\exp\!\left[-\ii\frac{\pi}{2}\hat n_b(\hat n_b-1)\right],
\end{equation}
and, with the pump initially in vacuum, the swap--Kerr--inverse-swap sequence induces the terminal operation
\begin{align}
 \hat U_p=\hat S_{\bm u}^{\dagger}\hat U_K\hat S_{\bm u}
 &=\exp\!\left[-\ii\frac{\pi}{2}\hat n_{\bm u}(\hat n_{\bm u}-1)\right] \nonumber \\
 &=\frac{e^{-\ii\pi/4}}{\sqrt2}
 \left[e^{\ii\pi\hat n_{\bm u}/2}
 +\ii e^{-\ii\pi\hat n_{\bm u}/2}\right],
 \label{S:Up}
\end{align}
where \(\hat n_{\bm u}=\hat c_{\bm u}^{\dagger}\hat c_{\bm u}\).
Here the first line follows from \(\hat{S}_{\bm u}^\dagger \hat n_{b} \hat{S}_{\bm u}=\hat n_{\bm u}\). To obtain the 
second line, we use the fact that the eigenvalues of $\hat{n}_{\bm u}$ are integers. So we can check the equation by replacing $\hat{n}_{\bm u}$ by integer values. Particularly, one can check separately for $\hat{n}_{\bm u} = 2m$, $\hat{n}_{\bm u} = 2m+1$, the equation in the second line is satisfied in both cases. 

It is useful to define the single-particle phase rotation
\begin{equation}
 V_{\bm u}(\alpha)
 =(I-\bm u\bm u^{\dagger})+e^{\ii\alpha}\bm u\bm u^{\dagger},
\end{equation}
for which
\begin{equation}
 e^{\ii\alpha\hat n_{\bm u}}\ket{N}_{\bm v}
 =\ket{N}_{V_{\bm u}(\alpha)\bm v}.
\end{equation}
Choosing
\begin{equation}
 \bm u_\star
 =\frac{\bm v_- -e^{\ii\chi}\bm v_+}{\sqrt2},
 \qquad
 \tilde{\bm v}=V_{\bm u_\star}(-\pi/2)\bm v_-,
 \label{S:programmedModes}
\end{equation}
gives
\begin{equation}
 \hat U_p\ket{N_{\tilde{\bm v}}}
 =\frac{e^{-\ii\pi/4}}{\sqrt2}
 \left(\ket{N_{\bm v_-}}+\ii e^{\ii N\chi}\ket{N_{\bm v_+}}\right),
\end{equation}
which is of the form in Eq.~\eqref{S:optimalCat}. The required input state is loaded from an externally prepared pump Fock state through
\begin{equation}
 \ket{0}_t\ket N_b
 \xrightarrow{\,\hat S_{\tilde{\bm v}}\,}
 (-\ii)^N\ket{N}_{\tilde{\bm v}}\ket0_b.
 \label{S:loading}
\end{equation}

The Kerr pulse is self-inverse because \(n(n-1)\) is even for every integer \(n\):
\begin{equation}
 \hat U_K^2=I,
 \qquad \hat U_p^2=I.
 \label{S:selfInverse}
\end{equation}
After interrogation, a calibrated terminal analysis operation corresponding to \(\hat U_{\theta_0}^{\dagger}(T)\) is applied, followed by the same \(\hat U_p\) and the inverse loading swap. At \(\theta=\theta_0\) this returns the state to \(\ket{0}_t\ket{N}_b\); away from the operating point the imperfect return carries the sensing signal. In the ideal fixed-\(N\) sector, distinguishing \(n_b=N\) from \(n_b\neq N\) therefore realizes the local return measurement described in the main text.

\section{Numerical details for the two-terminal benchmark}
\label{S:numerics}

We simulate the complete protocol for the differential signal
\begin{equation}
\hat H_{\rm sig}
=
\frac{\lambda}{2}(\hat n_1-\hat n_2),
\label{S:Hdiff}
\end{equation}
using the parameters quoted in the main text:
\(G/2\pi=2.05\,{\rm MHz}\),
\(|K|/2\pi=5.21\,{\rm MHz}\),
\(T_{1,T}=204\,\mu{\rm s}\),
\(T_{1,P}=16\,\mu{\rm s}\), and interrogation time
\(T=5\,\mu{\rm s}\).
The complete-swap and Kerr-pulse durations are
\(\tau_{\rm sw}=\pi/(2G)=0.122\,\mu{\rm s}\) and
\(\tau_K=\pi/|K|=0.096\,\mu{\rm s}\), respectively.

\subsection{Exact fixed-\texorpdfstring{$N$}{N} simulation and control sequence}

All coherent controls conserve the total excitation number
\(\hat N=\hat n_1+\hat n_2+\hat n_b\). We therefore work directly in the
fixed-\(N\) Hilbert space
\begin{equation}
\mathcal B_N=
\left\{
\ket{n_1,n_2,n_b}:n_1+n_2+n_b=N
\right\},
\qquad
D_N=\frac{(N+1)(N+2)}{2}.
\label{S:fixedNBasis}
\end{equation}
No oscillator cutoff is introduced. The initial state is
\(\ket{0,0,N}\), with all \(N\) excitations in the pump, and the
calculation is carried through \(N=100\).

For the differential signal, the two extremal sensing modes are the
physical terminals. With the phase convention used in the main text, the
nonlinear bright mode and the loading mode may be chosen as
\begin{equation}
\bm u=\frac{\bm e_2-\bm e_1}{\sqrt2},
\qquad
\tilde{\bm v}\sim
\frac{\bm e_1-\ii\bm e_2}{\sqrt2}.
\label{S:twoModes}
\end{equation}
For a normalized terminal mode \(\bm w\), define
\(\hat c_{\bm w}=\bm w^\dagger\hat{\bm a}\). The exchange Hamiltonian is
\begin{equation}
\hat H_{\rm sw}^{(\bm w)}
=
G\left(
\hat c_{\bm w}^{\dagger}\hat b+
\hat b^\dagger\hat c_{\bm w}
\right).
\label{S:HswapNum}
\end{equation}
Its Fock-basis matrix elements contain, for example,
\begin{equation}
\bra{\ldots,n_i+1,\ldots,n_b-1}
\hat H_{\rm sw}^{(\bm w)}
\ket{\ldots,n_i,\ldots,n_b}
=
G w_i\sqrt{(n_i+1)n_b},
\end{equation}
together with the Hermitian-conjugate process. Exchange Hamiltonians are
stored as sparse matrices in \(\mathcal B_N\), and their matrix
exponentials are applied directly to the state vector.

The nonlinear pulse is
\begin{equation}
\hat U_K=
\exp\!\left[
-\frac{\ii\pi}{2}
\hat n_b(\hat n_b-1)
\right],
\label{S:UKnum}
\end{equation}
and the loading and preparation sequence is
\begin{equation}
\hat S_{\tilde{\bm v}}^\dagger
\;\longrightarrow\;
\hat S_{\bm u}
\;\longrightarrow\;
\hat U_K
\;\longrightarrow\;
\hat S_{\bm u}^{\dagger}.
\label{S:prepSequence}
\end{equation}
The first pulse loads \(\ket{N_{\tilde{\bm v}}}\) from the pump, while
the last three pulses realize
\(\hat U_p=\hat S_{\bm u}^{\dagger}\hat U_K\hat S_{\bm u}\).
This process therefore takes
\begin{equation}
3\tau_{\rm sw}+\tau_K
=
0.462\,\mu{\rm s}.
\label{S:prepTime}
\end{equation}
After interrogation, the same nonlinear sequence is applied again,
followed by \(\hat S_{\tilde{\bm v}}\), so the decoding has the
same control duration. Preparation of the initial pump Fock state
\(\ket N_b\) is treated as an input resource and is not included in
this time analysis.

During interrogation all preparation controls are switched off. The
differential signal is diagonal in the physical Fock basis and gives the
exact evolution
\begin{equation}
\hat U_\lambda(T)
=
\exp\!\left[
-\frac{\ii\lambda T}{2}
(\hat n_1-\hat n_2)
\right].
\label{S:exactInterrogation}
\end{equation}
Thus a component \(\ket{n_1,n_2,n_b}\) acquires the phase
\(\exp[-\ii\lambda T(n_1-n_2)/2]\). No rotating-wave, or
short-time approximation is made to the sensing evolution.

\subsection{Ideal protocol and resource scaling}

We first verify the complete protocol in the absence of photon loss and
control errors. For the differential signal, the ideal probe has
\(F_Q=N^2\), and the same is obtained for the binary return measurement at $\lambda = 0$.
The diagonal terminal-number
covariances independently reproduce the same metrological resource through
the work--QFI relation discussed in the main text.

Figure~\ref{S:figIdealResources} summarizes these checks.
It shows the terminal QFI, the QFI reconstructed from terminal-number
correlations, and the binary return FI, each normalized by \(N^2\),
remain equal to unity through \(N=100\) within numerical precision. The
pump is also returned to vacuum after preparation, confirming that the
complete loading and nonlinear sequence produces the intended terminal
probe without residual pump occupation.

\begin{figure}[t]
\centering
\includegraphics[width=0.5\linewidth]{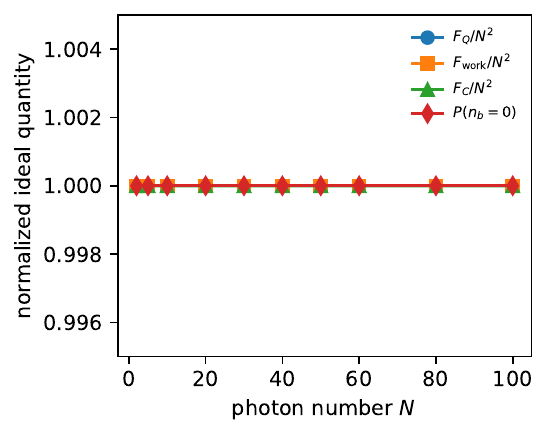}
\caption{Ideal resource accounting for the current protocol:
normalized terminal QFI, QFI reconstructed from terminal-number
correlations, binary return FI, and pump-vacuum probability after
preparation. }
\label{S:figIdealResources}
\end{figure}

\begin{figure}[t]
\centering
\includegraphics[width=0.90\linewidth]{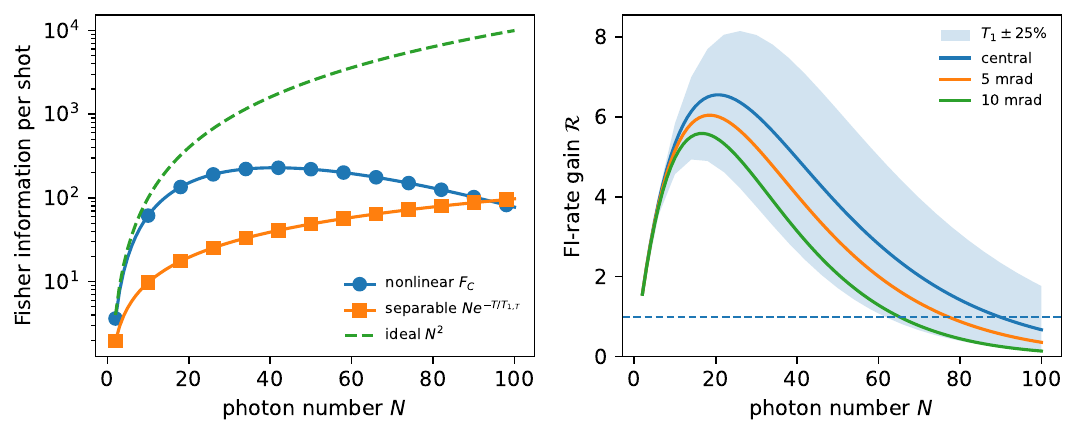}
\caption{Detailed two-terminal benchmark for the current protocol.
Left: binary Fisher information per shot with the central lifetimes,
together with the favorable lossy separable reference and ideal \(N^2\)
scaling. Right: Fisher-information-rate gain. The shaded band varies both
lifetimes by \(\pm25\%\); the additional curves include \(5\) and
\(10\,{\rm mrad}\) rms quasistatic differential-phase noise. The known
analysis bias is recalibrated for each \(N\).}
\label{S:figLossRate}
\end{figure}

\subsection{Local phase range and quasistatic phase noise}

Before including photon loss, it is useful to isolate the effect of the
finite phase range of the Heisenberg-limited probe. For ideal controls and
no loss, the return probability is
\begin{equation}
p_N(\phi)
=
\cos^2\!\left(\frac{N\phi}{2}\right),
\qquad
\phi=\lambda T .
\label{S:idealFringe}
\end{equation}
The corresponding binary Fisher information is \(F_C=N^2\) for the
accumulated phase \(\phi\), demonstrating Heisenberg scaling. At the same
time, the fringe width decreases as \(1/N\), so the range over which the
local response can be uniquely resolved becomes correspondingly narrower.

We next include a zero-mean quasistatic Gaussian fluctuation
\(\delta\) of the accumulated phase, with rms width \(\sigma_\phi\).
Averaging Eq.~\eqref{S:idealFringe} over \(\delta\) gives
\begin{equation}
\overline p_N(\phi)=\int
\frac{d\delta}{\sqrt{2\pi}\sigma_\phi}
e^{-\delta^2/(2\sigma_\phi^2)}
p_N(\phi+\delta)
=
\frac{1}{2}
\left[
1+
e^{-N^2\sigma_\phi^2/2}\cos(N\phi)
\right].
\label{S:idealPhaseAverage}
\end{equation}
The corresponding binary Fisher information is
\begin{equation}
F_C(\phi)
=
\frac{
N^2 e^{-N^2\sigma_\phi^2}\sin^2(N\phi)
}{
1-e^{-N^2\sigma_\phi^2}\cos^2(N\phi)
}.
\label{S:idealPhaseFINoise}
\end{equation}
It is maximized at the midpoint of any interference fringe,
\(\cos(N\phi)=0\), yielding
\begin{equation}
F_C^{\max}
=
N^2e^{-N^2\sigma_\phi^2}.
\label{S:idealPhaseFI}
\end{equation}
Thus quasistatic phase noise suppresses the ideal Heisenberg enhancement
through the dimensionless combination \(N\sigma_\phi\). Increasing \(N\)
is beneficial only while \(N\sigma_\phi\lesssim1\), illustrating the
tradeoff between the \(N^2\) sensitivity enhancement and the shrinking
phase range of the many-body interference fringe. See figure~\ref{S:figPhaseNoise}.

\begin{figure}[t]
\centering
\includegraphics[width=0.90\linewidth]{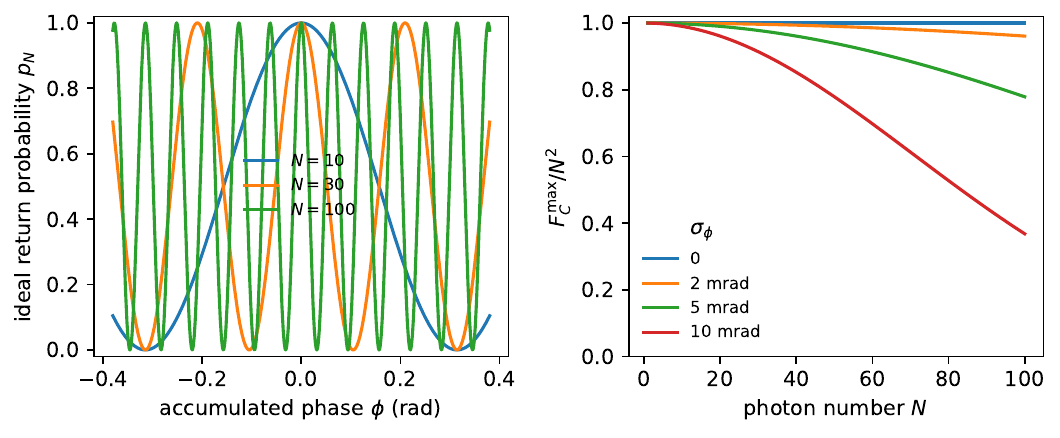}
\caption{Local phase range and quasistatic differential-phase noise.
Left: ideal return fringes for representative photon numbers.
Right: maximum binary Fisher information after Gaussian phase averaging,
normalized by \(N^2\).}
\label{S:figPhaseNoise}
\end{figure}

\subsection{Photon loss, return probability, and analysis bias}
Amplitude damping is included throughout loading, nonlinear preparation,
interrogation, and decoding. For the no-jump evolution, the Hamiltonian
of each active segment is replaced by
\begin{equation}
\hat H(t)\rightarrow
\hat H(t)
-\frac{\ii}{2}
\left[
\frac{\hat n_1+\hat n_2}{T_{1,T}}
+
\frac{\hat n_b}{T_{1,P}}
\right].
\label{S:HeffLoss}
\end{equation}
For the exchange pulses, the complete non-Hermitian sparse matrix is
exponentiated. During the Kerr pulse and differential interrogation,
both the coherent evolution and the damping are diagonal in the physical
Fock basis.

For the measured return event, this no-jump propagation is exact. Any
photon-loss jump lowers the total excitation number below \(N\), whereas
all subsequent controls conserve excitation number. A trajectory
containing a jump therefore cannot contribute to the final outcome
\(n_b=N\). Lower-excitation sectors need not be propagated for this
particular binary measurement.

At the end of the complete
loading--preparation--interrogation--decoding--unloading sequence, the
return probability is
\begin{equation}
p_N(\phi)
=
\left|
\braket{0,0,N|\psi_f(\phi)}
\right|^2,
\label{S:returnNumerical}
\end{equation}
where \(\ket{\psi_f(\phi)}\) is the unnormalized no-jump final state.
The two outcomes \(n_b=N\) and \(n_b\neq N\) give the binary Fisher
information
\begin{equation}
F_C(\phi)
=
\frac{\left[\partial_\phi p_N(\phi)\right]^2}
{p_N(\phi)\left[1-p_N(\phi)\right]}.
\label{S:binaryFI}
\end{equation}

Photon loss qualitatively changes the behavior at the return maximum.
In the ideal lossless protocol, 
Eq.~\eqref{S:binaryFI} at \(\phi=0\) has the finite limit \(F_C=N^2\).
With loss, however, \(p_N(0)<1\) while symmetry still gives
\(\partial_\phi p_N|_{\phi=0}=0\). Consequently, this particular binary
measurement has \(F_C(0)=0\), and its most sensitive local operating
point occurs at a nonzero accumulated phase.

For each \(N\), we therefore evaluate the maximum of \(F_C(\phi)\) over
one equivalent fringe,
\begin{equation}
0<\phi<\frac{\pi}{N}.
\label{S:operatingRange}
\end{equation}
This simply selects the local operating point of the physical response
\(p_N(\phi)\); no additional phase variable or auxiliary bias is
introduced. For the central parameters and zero phase noise, the maximum
FI-rate gain occurs at \(N=21\), where the optimal accumulated phase is
\begin{equation}
\phi=0.14871885\,{\rm rad}.
\end{equation}
For the peak points of the \(5\) and \(10\,{\rm mrad}\) quasistatic
phase-noise curves, the corresponding operating phases are
\(0.14808632\,{\rm rad}\) at \(N=18\) and
\(0.14666930\,{\rm rad}\) at \(N=17\), respectively.

\subsection{Realistic Fisher-information-rate benchmark}

Because preparation and decoding require finite time, we compare the
Fisher information accumulated per unit experimental time. One nonlinear
sensing cycle has duration
\begin{equation}
\tau_{\rm cyc}
=
2(3\tau_{\rm sw}+\tau_K)+T,
\end{equation}
and therefore
\begin{equation}
\widetilde F_C
=
\frac{F_C}
{2(3\tau_{\rm sw}+\tau_K)+T}.
\label{S:Fcrate}
\end{equation}
As a favorable separable reference, we take a single photon prepared in
the optimal two-terminal Ramsey state and assign it zero preparation and
readout overhead. Equal terminal loss is then a phase-independent erasure,
giving
\begin{equation}
\widetilde F_{\rm sep}
=
\frac{e^{-T/T_{1,T}}}{T}.
\label{S:Fseprate}
\end{equation}
For a total resource of \(N\) photons, the rate gain is
\begin{equation}
\mathcal R
=
\frac{\widetilde F_C}
{N\widetilde F_{\rm sep}}.
\label{S:rateGain}
\end{equation}
Thus \(\mathcal R>1\) means that the nonlinear protocol acquires more
Fisher information per unit experimental time even after including its
loading, nonlinear preparation, and decoding overhead. The numerical
\(F_C\) is evaluated with respect to the accumulated differential phase
\(\lambda T\). Multiplying both the nonlinear and separable Fisher
informations by \(T^2\) converts them to Fisher information for the
frequency shift \(\lambda\), leaving \(\mathcal R\) unchanged.

To test lifetime uncertainty, both lifetimes are varied together as
\begin{equation}
(T_{1,T},T_{1,P})
\longrightarrow
0.75(T_{1,T},T_{1,P})
\quad\text{and}\quad
1.25(T_{1,T},T_{1,P}),
\end{equation}
while \(G\), \(K\), and \(T\) are held fixed. For quasistatic phase
noise, the lossy return probability is first averaged according to
\begin{equation}
\overline p_N(\phi)
=
\int
\frac{d\delta}{\sqrt{2\pi}\sigma_\phi}
e^{-\delta^2/(2\sigma_\phi^2)}
p_N(\phi+\delta),
\label{S:phaseAverage}
\end{equation}
after which the binary Fisher information is evaluated from
\(\overline p_N\). The integral is calculated using ten-point
Gauss--Hermite quadrature, and the known bias \(\phi_b\) is recalibrated
for every \(N\) and noise strength.

Figure~\ref{S:figLossRate} summarizes the realistic two-terminal
performance. The left panel shows the binary Fisher information per shot.
At small \(N\), the nonlinear sensor follows the ideal \(N^2\) growth,
while at larger \(N\) photon loss increasingly suppresses the probability
of completing the full protocol without a jump and eventually causes the
Fisher information to decrease. The favorable separable reference, by
contrast, grows only linearly with \(N\). This directly illustrates the
competition between coherent many-body enhancement and the increasing
fragility of larger Fock-state superpositions.

The right panel of Fig.~\ref{S:figLossRate} shows the corresponding
FI-rate gain and reproduces the main-text result. For the central
lifetimes and zero phase noise, the gain reaches
\(\mathcal R\simeq6.55\) at \(N=21\). Varying both lifetimes by
\(\pm25\%\) gives the shaded uncertainty band. Quasistatic differential
phase noise reduces the gain and moves the optimum toward smaller photon
numbers: the peak becomes approximately \(6.04\) at \(N=18\) for
\(5\,{\rm mrad}\) rms noise and \(5.58\) at \(N=17\) for
\(10\,{\rm mrad}\). The finite optimal \(N\) therefore results from the
competition between the ideal \(N^2\) enhancement, photon loss, finite
control time, and phase uncertainty.

\subsection{Higher-order pump nonlinearity}

The ideal nonlinear pulse assumes that the leading pump nonlinearity is
the self-Kerr term. To quantify sensitivity to a higher-order correction,
we add
\begin{equation}
\hat H_6
=
\frac{K_6}{6}
\hat n_b(\hat n_b-1)(\hat n_b-2).
\label{S:H6}
\end{equation}
This term is included during the Kerr pulse of the same preparation
sequence used above. For each \(K_6/K\) and \(N\), the Kerr-pulse duration
is recalibrated near its ideal value to maximize the prepared-state QFI;
no compensating terminal rotation or additional phase correction is
introduced.

Figure~\ref{S:figHigherOrder} shows the residual metrological
infidelity after this recalibration. At \(N=100\), the QFI ratios are
approximately \(0.9986\), \(0.9873\), and \(0.8804\) for
\(K_6/K=10^{-4}\), \(3\times10^{-4}\), and \(10^{-3}\), respectively.
The error grows with both \(N\) and \(K_6/K\). Recalibrating the pulse
duration removes the dominant shift of the nonlinear phase, but cannot
eliminate the different occupation dependence introduced by
\(\hat H_6\). Higher-order pump nonlinearities therefore become an
increasingly important hardware constraint as the photon number is
increased.

\begin{figure}[t]
\centering
\includegraphics[width=0.68\linewidth]{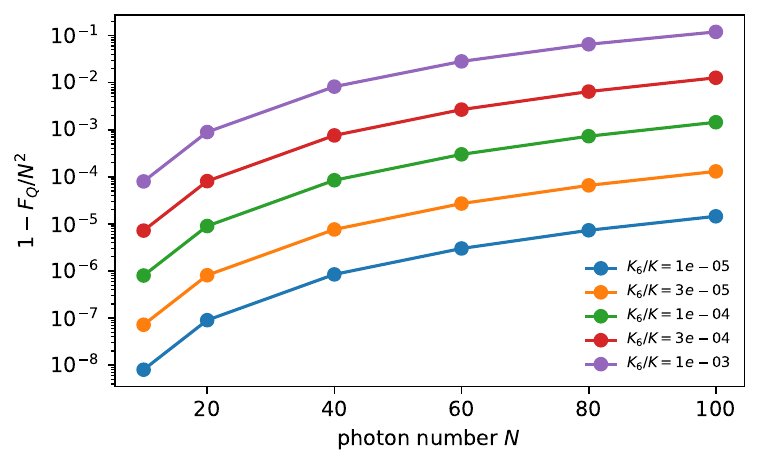}
\caption{Sensitivity to the leading higher-order pump nonlinearity in
the current protocol. The Kerr-pulse duration is recalibrated for each
\(K_6/K\) and \(N\), but no additional terminal phase correction is
applied.}
\label{S:figHigherOrder}
\end{figure}

\subsection{Numerical convergence}

The derivative entering the lossy binary Fisher information is evaluated
using the centered finite difference
\begin{equation}
\partial_\phi p_N
\simeq
\frac{p_N(\phi+h)-p_N(\phi-h)}{2h},
\qquad
h=\frac{2\times10^{-5}}{N}.
\label{S:finiteDifference}
\end{equation}
Repeating the current-protocol calculation at \(N=20,50,\) and \(100\)
with steps \(h/2\), \(2h\), and \(4h\) changes the resulting Fisher
information by less than \(2.1\times10^{-9}\) relatively. The exact
fixed-\(N\) basis eliminates bosonic-cutoff error, while the remaining
exchange propagators are evaluated through sparse action of the matrix
exponential. The Kerr pulse, differential interrogation, and their
associated loss factors are diagonal in the physical Fock basis and are
applied directly.

\section{Three-terminal example and control robustness}
\label{S:threeTerminal}

To show that the programmable bright-mode rule is not restricted to a
pairwise contrast, consider the exact finite-time generator
\begin{equation}
K_T=\begin{pmatrix}
1/5&-4/5&-2/5\\
-4/5&1/5&-2/5\\
-2/5&-2/5&-1/5
\end{pmatrix}.
\label{S:Kthree}
\end{equation}
Its extremal eigenpairs are
\begin{equation}
\kappa_-=-1,\quad \bm v_-=(1,1,1)^T/\sqrt3,
\qquad
\kappa_+=1,\quad \bm v_+=(1,-1,0)^T/\sqrt2.
\end{equation}
For \(\chi=0\), the programmed nonlinear bright mode is
\begin{equation}
\bm u_\star=\frac{\bm v_--\bm v_+}{\sqrt2}
=\begin{pmatrix}
1/\sqrt6-1/2\\
1/\sqrt6+1/2\\
1/\sqrt6
\end{pmatrix},
\label{S:uThree}
\end{equation}
which has support on all three terminals. The loading mode is
\(\tilde{\bm v}=V_{\bm u_\star}(-\pi/2)\bm v_-\). Since the many-body
generator has endpoints \(-N\) and \(+N\), the ideal probe obeys
\begin{equation}
F_Q^{\max}=4N^2.
\label{S:FQthree}
\end{equation}

We propagate the current
loading--swap--Kerr--inverse-swap sequence exactly in the fixed-\(N\)
space spanned by \(\ket{n_-,n_+,n_b}\), where \(n_-+n_++n_b=N\).
The dimension is again \((N+1)(N+2)/2\), and no oscillator cutoff is
introduced. In this basis the sensing generator is
\(\hat n_+-\hat n_-\), so the prepared-state QFI is evaluated directly as
\(F_Q=4\Var(\hat n_+-\hat n_-)\).

To quantify coherent preparation errors, we vary one control at a time. A
swap-area error replaces every complete-swap area \(\pi/2\) by
\((\pi/2)(1+\epsilon_{\rm sw})\); a Kerr-area error replaces \(\pi\) by
\(\pi(1+\epsilon_K)\); and a bright-mode error rotates \(\bm u_\star\)
by a small angle toward the orthogonal bisector
\((\bm v_-+\bm v_+)/\sqrt2\). Figure~\ref{S:figThreePrep} shows
representative \(1\%\) swap-area, \(0.1\%\) Kerr-area, and
\(10\,\mathrm{mrad}\) bright-mode errors. At \(N=50\) they retain,
respectively,
\begin{equation}
\frac{F_Q}{4N^2}=0.9990,\qquad 0.9941,\qquad 0.9996.
\end{equation}

The trends in Fig.~\ref{S:figThreePrep} distinguish the three error
mechanisms. The swap-area and bright-mode errors produce nearly
\(N\)-independent infidelities of order \(10^{-3}\) and \(4\times10^{-4}\),
respectively, over the range shown. The Kerr-area error is much smaller at
low \(N\) but grows steadily with photon number and becomes the dominant
preparation error beyond \(N\sim20\). This enhanced sensitivity is expected
because a Kerr-area error produces occupation-dependent phase errors that
accumulate more strongly across a large-\(N\) superposition.

\begin{figure}[t]
\centering
\includegraphics[width=0.65\linewidth]{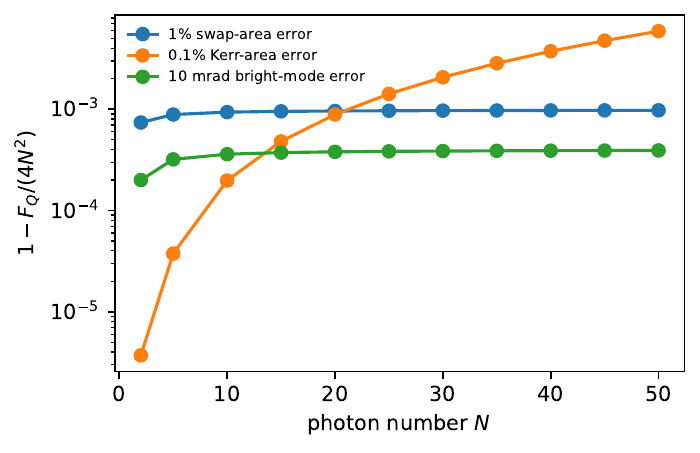}
\caption{Three-terminal preparation robustness for the current protocol.
Shown is the metrological infidelity for representative coherent control
errors applied one at a time.}
\label{S:figThreePrep}
\end{figure}

We separately test decoder mismatch by preparing the optimal probe ideally
and applying the same three errors only during decoding. The return
probability is converted to the binary Fisher information, with the known
analysis bias recalibrated over \(0<\phi_b<\pi/N\) for each \(N\). At
\(N=50\), the representative swap, Kerr, and bright-mode mismatches give
\begin{equation}
\frac{F_C}{4N^2}=0.9730,\qquad 0.9280,\qquad 0.9950,
\end{equation}
respectively. The Kerr area is therefore the most demanding of these three
coherent calibrations as \(N\) increases.

Figure~\ref{S:figThreeReadout} also shows that decoder mismatch is more
restrictive than preparation mismatch for the same nominal control errors.
The \(10\,\mathrm{mrad}\) bright-mode mismatch remains weak, retaining more
than \(99.5\%\) of the ideal FI at \(N=50\). A \(1\%\) swap-area mismatch
produces a gradual reduction to about \(97.3\%\), whereas the \(0.1\%\)
Kerr-area mismatch falls more rapidly at large \(N\) and retains about
\(92.8\%\) at \(N=50\). The readout therefore places the tightest coherent
calibration requirement on the nonlinear phase area in this example.

\begin{figure}[t]
\centering
\includegraphics[width=0.65\linewidth]{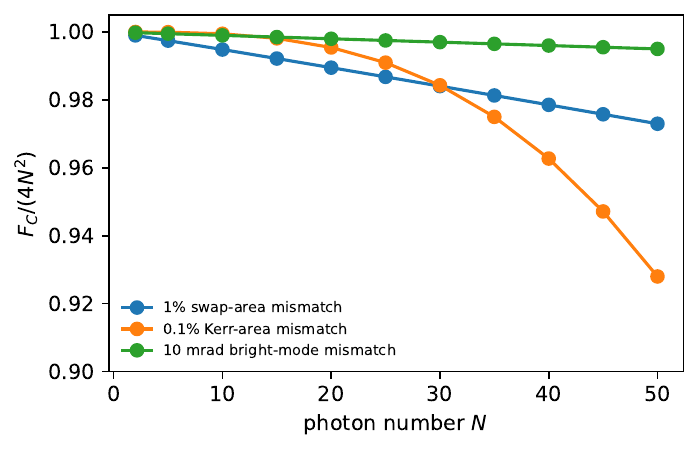}
\caption{Three-terminal readout robustness. The probe is prepared ideally
and the indicated mismatch is applied only to the decoder. The known
analysis bias is recalibrated for each \(N\).}
\label{S:figThreeReadout}
\end{figure}

\section{Physical terminal-work correspondence}
\label{S:work}

For the diagonal signal
\begin{equation}
 \hat H_{\rm sig}(\bm\theta)=\sum_i\theta_i\hat n_i,
\end{equation}
the generators are \(T\hat n_i\), and a pure prepared probe has
\begin{equation}
 (F_Q)_{ij}=4T^2\Cov(\hat n_i,\hat n_j).
 \label{S:QFImatrix}
\end{equation}
Let \(\hat U_{\rm prep}\) denote the complete state-preparation operation, including loading from the pump. The physical terminal-energy change is
\begin{equation}
 \hat W_i
 =\Omega_i\left(\hat U_{\rm prep}^{\dagger}\hat n_i\hat U_{\rm prep}-\hat n_i\right).
 \label{S:terminalWork}
\end{equation}
If the terminals begin in vacuum, the initial occupations are sharp and vanish, so the final energy changes are simply \(W_i=\Omega_i n_i\) in a number-resolved measurement. Consequently,
\begin{equation}
 \Cov(W_i,W_j)
 =\Omega_i\Omega_j\Cov(n_i,n_j),
\end{equation}
and Eq.~\eqref{S:QFImatrix} becomes
\begin{equation}
 (F_Q)_{ij}
 =\frac{4T^2}{\Omega_i\Omega_j}\Cov(W_i,W_j).
 \label{S:workQFI}
\end{equation}
Thus the complete QFI matrix for diagonal frequency shifts can be certified from terminal number statistics immediately after preparation, without applying the unknown signal. For a single parameter
\(\hat H_{\rm sig}=\theta\sum_iq_i\hat n_i\), the scalar QFI is obtained by projection,
\begin{equation}
 F_Q^{(\theta)}=\bm q^T\mathbf F_Q\bm q
 =4T^2\Var\!\left(\sum_iq_i\hat n_i\right).
\end{equation}
The work interpretation in this section is special to signals diagonal in the physical terminal basis; it is not used for the general off-diagonal sensing protocol.
\end{widetext}
\end{document}